\documentclass{raa} 
\usepackage{graphicx,times}
\usepackage{epsfig}
\usepackage{longtable}
\usepackage{setspace}
\usepackage{natbib}

\begin{document}

\title{The kilo-second variability of  X-ray sources in nearby galaxies}
\volnopage{Vol.0 (200x) No. 0, 000--000}
\setcounter{page}{1}
\author{Soma Mandal \inst{1} \and Ranjeev Misra \inst{2} \and Gulab C. Dewangan \inst{2}}
\institute{Department of Physics, Taki Government College, West Bengal, India; {\it soma@iucaa.ernet.in}\\
\and 
 Inter-University Centre for Astronomy and
Astrophysics,  Post Bag 4, Ganeshkhind, Pune-411007,\\ India$~$ ;  {\it rmisra@iucaa.ernet.in ; gulabd@iucaa.ernet.in} }

\date{Received~~....; accepted~~....}

\abstract
{{\it Chandra} observations of 17 nearby galaxies were
analysed and 166 bright sources with X-ray counts $> 100$, were chosen
for  temporal analysis. Fractional root mean square variability amplitudes were
estimated for lightcurves binned at $\sim 4$ ksec and of length
$\sim 40$ ksec. While there are nine ultra-luminous X-ray
sources (ULXs) with unabsorbed luminosity (in 0.3-8.0 keV band) $L > 10^{39}$ erg s$^{-1}$ in
the sample for which the fractional r.m.s variability is constrained
to be $< 10$\%, only two of them show variability. One of the variable ULXs exhibits a secular transition
and has a ultra-soft spectrum with temperature $\sim 0.3$ keV while the other is a rapidly
varying source in NGC 0628, which
has been previously compared to the Galactic micro-quasar GRS
1915+105. These results seem to indicate that ULXs are typically
not highly variable in ksec time-scales, except for some ultra-soft
ones. Among the relatively low luminosity sources
($L \sim 10^{38}$ erg s$^{-1}$ ) we find five of them to be variable. Apart from an earlier known source in NGC 1569, we
identify a source in NGC 2403, which exhibits persistent high amplitude fluctuations.
The variability of the sources in general, do not seem to be correlated with hardness,
which indicates that they may not be due to variations in any absorbing material, but
instead could reflect inner accretion disk instabilities.
}

\authorrunning{Mandal, Misra \& Dewangan}
\titlerunning{The Kilo-second variability of X-ray sources}

\maketitle
\section{Introduction} 

The unprecedented angular resolution of {\it Chandra} allows for the detailed
study of X-ray point sources in nearby galaxies. Most of these sources are
expected to be X-ray binaries harbouring neutron stars or black holes, similar to
the ones found in our Galaxy. Some of these sources  have an isotropic luminosity exceeding   $> 10^{39}$ erg s$^{-1}$
and are called Ultra-Luminous X-ray sources (ULXs). Since
these sources emit radiation at a rate larger than the Eddington
limit for a ten-solar mass black hole, they may be harbouring 
black holes of mass $10 \, M_\odot\! < \! M \! <\! 10^5
\,M_\odot$ \citep{Col99,Mak00} where the upper limit 
is constrained by the argument that a
more massive black hole would have settled into the nucleus due to
dynamical friction \citep{Kaa01}. 
In this interpretation, these black holes have mass in the intermediate mass range between those of
stellar mass black holes found in Galactic X-ray binaries and those associated
with  Active Galactic Nuclei (AGN) and hence are called Intermediate Mass Black Holes (IMBH).
Alternatively, if these sources are powered by 
stellar mass black holes which are radiating at a super Eddington level \citep{Beg02, Kin08}, this would 
imply that their accretion state is significantly different than that of AGN and X-ray binaries.
The process by which such black holes are created \citep{Tan00,Mad01,Por02} and their environment which allows
such sustained high accretion rates \citep{Kin01} are largely unknown and understanding them may lead to  radical 
shifts in the present paradigms of stellar and binary evolution and the history of the Universe.
For a review of their observational properties and implications see \citet{Mil04,Mil05,Sor11}.  

While there have been several extensive analysis of individual ULX \citep[e.g.,][]{Fen05,Fen06,Mil04,Dev08,Dew05,Dew06b,Agr06},
it is also important to undertake systematic spectral and temporal analysis of ULXs and other lower luminosity sources
to bring out any systematic differences between the two. Lower luminosity sources ($L_X \sim 10^{38}$ erg s$^{-1}$) are expected
to be similar to typical Galactic black hole systems. Galactic black hole systems show variability on short time-scales
of $\sim 10$ Hz \citep[e.g.,][]{Now99} and long term ($\sim$ months) variations which are usually associated with spectral state
transitions \citep[e.g.,][for a review]{Zdz04,Rem06}. In the \textquotedblleft Hard \textquotedblright state, the spectrum is dominated by a power-law,
often modelled to be thermal Comptonization, while in the \textquotedblleft Soft \textquotedblright state, it is a thermal one arising from 
a multi-colour disk. X-ray binaries show long term ($> $  days) variability as super-orbital modulations
\citep{Kot12} and aperiodic behaviour \citep{Gil05}.
On time-scales of hours, Galactic black hole systems typically do not seem to show persistent large variability, with
the exceptions of the micro-quasar GRS 1915+105 \citep{Mir94} and the recently discovered source IGR J17091-3624 \citep{Alt11}.
Occasional flares or large amplitude variations on minute timescales have been reported for some
systems like V4641 Sgr \citep{Mai06}.

Systematic spectral studies of a sample of sources have also been undertaken to uncover
differences between ULXs and less luminous sources. 
\citet{Swa04} fitted the spectra of such a large sample
with an absorbed power-law model and found no bimodal distribution on the
spectral index. However, using a disk black body model, a bimodal distribution 
was revealed at least for very high luminosity ($ > 10^{40}$ erg s$^{-1}$) sources in the samples obtained from 
{\it XMM-Newton} \citep{Win06} and {\it Chandra} \citep{Dev07}. One set of high luminosity
sources have  temperatures $kT \sim 0.1$ keV while the other have systematically
higher temperatures $kT \sim 1 $ keV. It is
not clear whether they represent two kinds of sources or are two spectral states of a generic
type. Such a bi-modal distribution is not evident for low luminosity sources ($\sim 10^{38}$ erg s$^{-1}$).
The hard sources are dominated by an optically thick Comptonized component
while the soft ones are compatible with being emission 
from accretion disks around IMBH accreting matter at $\sim 0.1$ of the Eddington limit\citep{Dev07}. However, \citet{Gla09b}
argue that detailed X-ray spectral analysis of bright sources do not favour an IMBH interpretation.

An important aspect of ULXs is that they are generally variable on a wide range of time-scales.
For a review of the time variability of X-ray sources in nearby galaxies see \citet{Fab06}. Nearly half
of the X-ray sources in M31 \citep{Kon02}, M82 \citep{Chi11}, the Antennae galaxies \citep{Zez06}, NGC 6946, NGC 4485/4495 
\citep{Fri08} are variable on
time-scales of weeks to  months. ULXs seem to be more variable as 12 of 14 sources in the Antennae galaxies
show long term variability. These long term variability are sometimes associated with spectral
state changes \citep{Col99,Kub01,Dew04,Fen06,Fen09,Glad9,Kajava09,Dew10}. While rapid variability (on $\sim 100$ sec time-scales) are
difficult to detect in low count rate sources,  quasi-periodic oscillations have been
reported from the bright ULXs -- M~82~X-1 \citep{Str03,Dew06a,Fen10} at $54$ mHz and $114$ mHz, Holmberg~IX~X-1
at $202$ mHz \citep{Dew06b} and NGC~5408~X-1 at $20$ mHz \citep{Str07,Str09}. These results provide strong evidence that the ULXs are not distant
background AGN and the frequencies of the oscillations suggest that they harbour IMBH although \cite{Mid11}
also argue for a super-Eddington interpretation. \citet{Hei09} studied the variability in the frequency range, $10^{-3}$-$1$ Hz, for
16 ULXs using {\it XMM-Newton} observations. The found that while six sources show intrinsic variability, more interestingly
they could could constrain the variability of four sources to be significantly less than what is expected from black hole
binaries and AGN.
Like X-ray binaries, different spectral states of ULXs should be associated with different temporal
behaviour and this was shown to be the case for the bright ULX in NGC 1313, which was variable 
in the high flux state and not in the low one \citep{Dew10}. A bright source in NGC 628,
is known to have rapid variability in short time-scales of $1000$ s \citep{Kra05}. This source is of particular
interest because it maybe an extragalactic example of the Galactic black hole system
GRS 1915+105. Large amplitude variability in ksec time-scales
have also been reported from two X-ray sources in NGC 1569 \citep{Hei03}. 
\begin{center}

\centering
\begin{table}
\caption{Sample galaxy properties}
~\\
\begin{tabular}{lccccc}
\hline
Galaxy & Distance$(Mpc)$ & ObsID & {$T_{exp} (ks)$} & {$N(\ge100)$}\\
\hline
NGC0628 & $9.7$ & $2058$ &$46.16$ & $5$ \\  
NGC0891 & $10.0$ & $794$ & $50.90$ & $12$ \\
NGC1291 & $8.9$ & $795$ & $39.16$ & $6$ \\   
NGC1399 & $18.3$ & $319$ & $55.94$ & $27$ \\
NGC1569 & $2.2$ & $782$ & $96.75$ & $11$ \\ 
NGC2403 & $3.1$ & $2014$ &$35.59$ & $2$ \\   
NGC3556 & $14.1$ & $2025$ & $59.36$ & $14$ \\ 
NGC3628 & $10.0$ & $2039$ & $57.96$ & $9$ \\ 
NGC4125 & $24.2$ & $2071$ & $64.23$ & $6$ \\ 
NGC4365 & $20.9$ & $2015$ & $40.42$ & $3$ \\  
NGC4579 & $21.0$ & $807$ & $33.90$ & $3$ \\ 
NGC4631 & $7.6$ & $797$ & $59.21$ & $6$ \\  
NGC4649 & $16.6$ & $785$ & $36.87$ & $13$ \\ 
NGC4697 & $11.8$ & $784$ & $39.25$ & $8$ \\  
NGC5128 & $4.0$ & $962$ & $36.50$ & $17$ \\ 
NGC5775 & $26.7$ & $2940$ & $58.21$ & $12$ \\
NGC6946 & $5.5$ & $1043$ & $58.28$ & $12$ \\
\hline
\end{tabular}
\tablecomments{0.7\textwidth}{ $T_{exp}$, exposure time in ks;
N, number of point sources with net counts $\ge$ 100}
\label{Tab_list} 
\end{table}
\end{center}

The main aim of this work is to study and quantify the variability of a sample of {\it Chandra} detected X-ray sources
in nearby galaxies in the time-scales of $\sim 10$ ksec. Given the faintness of these sources and {\it Chandra's} sensitivity we
expect to detect only large amplitude variations. On these time-scales, such large variability may be due to
spectral state transitions and this study will help us identify sources which have undergone
such transitions. While such transitions maybe rare events, a 'snapshot' study of a large enough sample, may shed light on how often
they occur on these time-scales i.e. the number of such sources may provide an idea of the duty cycle of such events. On the other
hand, the large amplitude variability could reflect  persistent aperiodic/periodic behaviour of the source. Finally, the study may reveal differences  
between regular X-ray sources and ULXs in terms of their k-sec variability properties. 

\section{Observations and Data analysis}

\cite{Dev07} have  analysed thirty galaxies observed by {\it Chandra}. They
selected those sources with counts $> 60$ and whose  spectra were not contaminated
by excessive diffuse emission and not affected by photon pile up. For the timing analysis, we
have chosen 17 of these galaxies
that have an exposure time roughly equal to or greater than $40$ ksec. 
In order to obtain reasonable statistics we have limited our analysis to sources
having net counts $> 100$. 
The names of these galaxies, the distances to them, the {\it Chandra} observation IDs and the
number of sources used in this analysis  are listed in Table \ref{Tab_list}. A total
of 166 sources were chosen for temporal analysis.

The estimation of the unabsorbed luminosity of a source, in general can depend
on the specific spectral model used. Hence, \citet{Dev07} fitted each source
with two models an absorbed power-law  and an absorbed disk black body and
have tabulated the best fit spectral parameters.  For
some sources, one of the models provides a significantly better fit (with a C$_{\it{stat}}$ difference greater than 2) and the unabsorbed luminosity of the source can then be estimated using the
better fitting model. However, for extreme soft spectra which give a best photon index $\Gamma > 4$, we choose the
disk black body representation, irrespective of the spectral fitting statistics.  For cases when both spectral models provide similar fits to the
data, we conservatively choose the lower of the unabsorbed luminosities estimated from
the models. Through out this work, unless otherwise specified, we have used the above
criterion to estimate the unabsorbed luminosity of the sources.

For each source, background subtracted  light curves were generated
with a time bin size of $4$ ksec by using CIAO tool dmextract. Following \citet{Vau03}, we calculate the
fractional root mean square variability amplitude, $F_{var}$ of the light curves.
The normalised variance of a light curve, $x_i$ is
\begin{equation}
 \sigma^2_{N} = \frac{1}{(N-1) {\bar x}^2}\sum^{N}_{i=1} (x_i - \bar x )^2
\end{equation} 
where $N$ is the number of points of the light curve and $\bar x$ is the mean.
A fraction of the variance is due to measurement errors and hence it is
convenient to define normalised excess variance as
\begin{equation}
\sigma_{NXS}^2 = \sigma^2_{N} - \bar {\sigma_{err}^2}
\end{equation}
where $\bar {\sigma_{err}^2} = \frac{1}{N{\bar x}^2  }\sum^N_{i=1} \sigma^2_{err,i}$ is the
normalised mean square error. Finally the fractional amplitude is defined as
$F_{var} = \sqrt{{\sigma_{NXS}^2}}$. Using simulation, \citet{Vau03} have estimated the
error on $\sigma^2_{NXS}$ to be
\begin{equation}
\hbox{err}(\sigma^2_{NXS})^2 = (\sqrt\frac{2}{N} \sigma^2_{err})^2 + ( \sqrt\frac{\sigma^2_{err}}{N} 2 F_{var})^2
\end{equation}

At the n-sigma level, a source is considered to be variable only if 
$\sigma^2_{NXS} > n  \hbox{err}(\sigma^2_{NXS})$. For such variable sources, 
the n-sigma upper and lower limits on $F_{var}$ can be estimated to be \\
$F_{var, u} = \sqrt{\sigma^2_{NXS} + n\hbox{err}(\sigma^2_{NXS})}$ and  
$F_{var, l} = \sqrt{\sigma^2_{NXS} - n\hbox{err}(\sigma^2_{NXS})}$ respectively.
For non-variable sources, $F_{var, u}$ provides an upper limit on any variability
that may be hidden in the measurement errors.

To measure the variability of the sources in a similar range of time-scales, we
bin all the data in $\sim 4$ ksec time bins and take the length of the the light curves
to be $\sim 40$ ksec i.e. $N \sim 10$.  For NGC 1569, the long exposure time 
of $\sim 90$ ksec, allowed us to obtain two light curves of $\sim 40$ ksec. The
fractional variability was averaged over the two light curves for this case. 
Using a $\sim 4$ ksec time bin, for most sources we place a 2-sigma
upper limit on  $F_{var} < 0.4$.  Choosing a smaller time bin would result in
a large number of sources having upper limits of $ F_{var}$  $\sim 0.8$, which is not
constraining. A larger time bin would not allow for many sources to have at least
ten data points.

\section{Results}

\begin{figure*}
\centering
\begin{minipage}{170mm}
\centering
\includegraphics[width=\textwidth, angle=0]{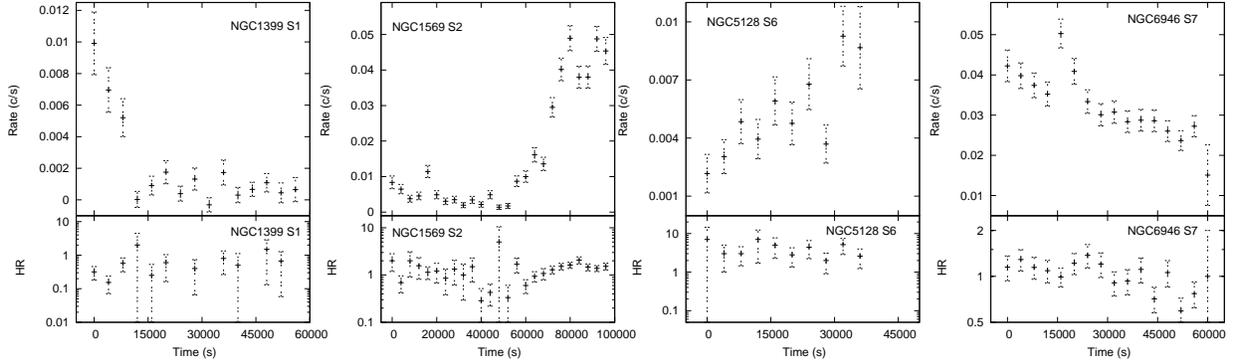}
\caption{{\small The light curves and the hardness ratio (i.e. count ratio in the
energy bands $0.3$-$1$ and $1$-$8$ keV bands) for the four sources that show secular
transitions or flares. The spectral properties of these sources are tabulated in
Table $2$. The source in NGC 5128 does not show such
variability in other observations and hence for this source flares/transitions are not persistent. }}
\label{LC_sec}
\end{minipage}
\end{figure*}

\begin{figure}[ht]
\begin{minipage}[t]{0.495\linewidth}
\centering
\includegraphics[height=110mm,angle=0]{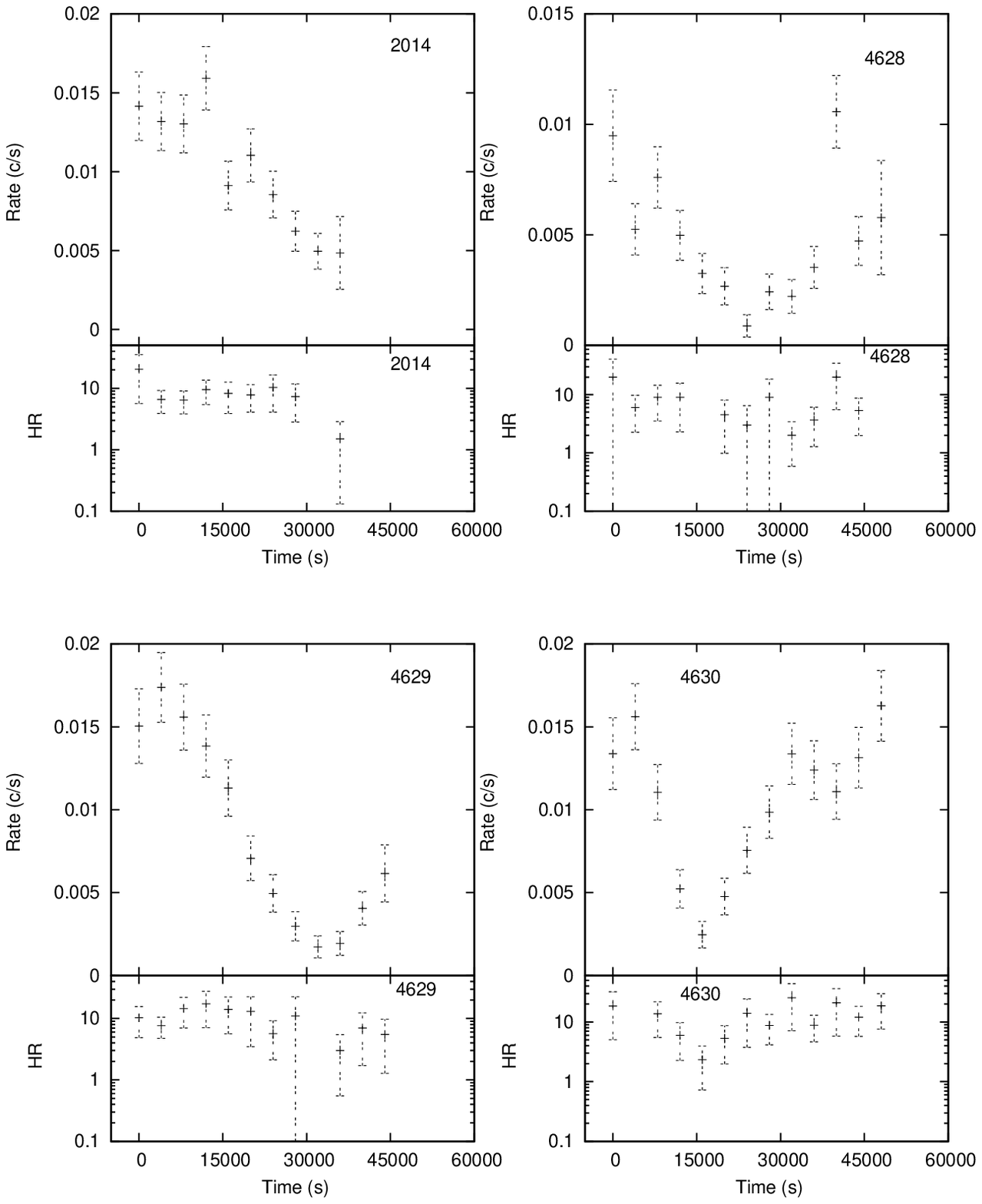}
\caption{The light curves and the hardness ratio (i.e. count ratio in the
energy bands $0.3$-$1$ and $1$-$8$ keV bands) for a source in NGC 2403, S4.
The panels are for four {\it Chandra} observations and are labelled by the
observation ID numbers. This source shows persistent flare like variability. }
\label{ngc2403}
\end{minipage}
\begin{minipage}[t]{0.495\linewidth}
\centering
\includegraphics[width=55mm,height=120mm,angle=0]{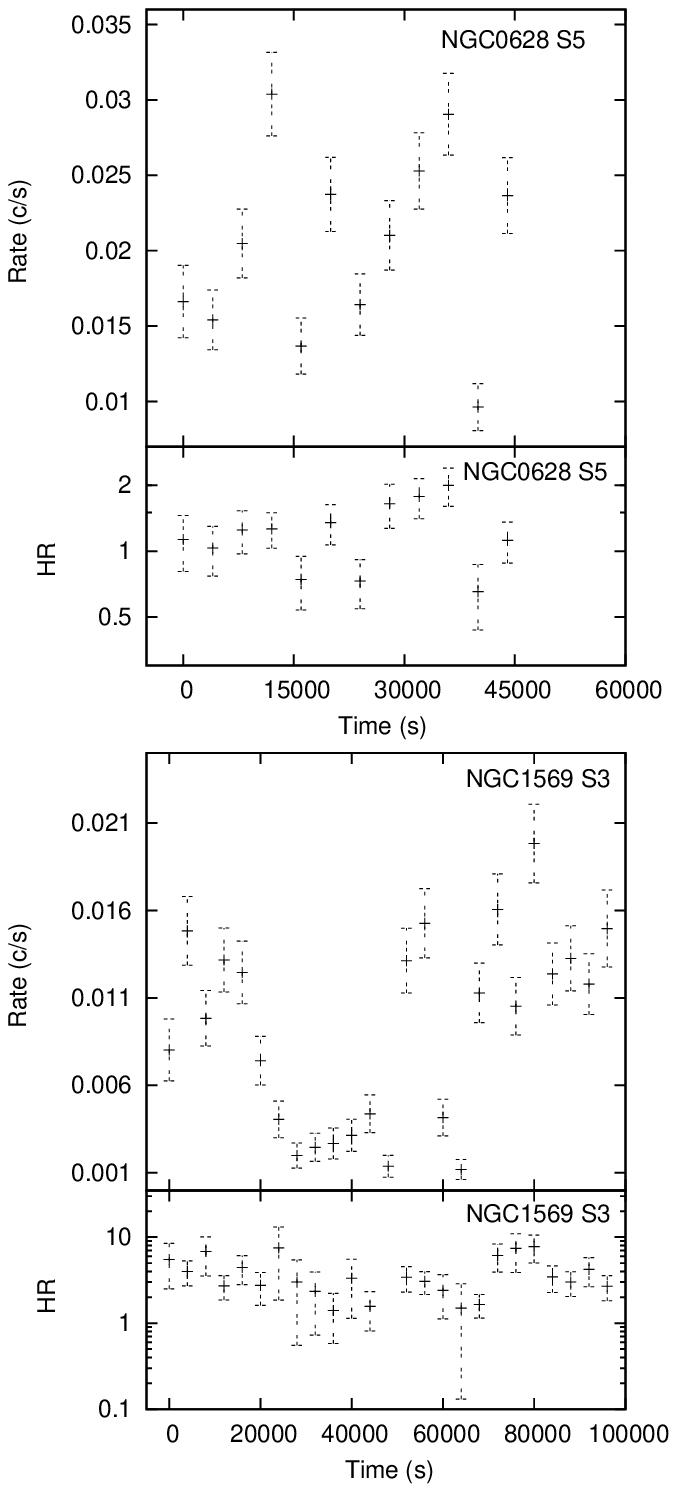}
\caption{The light curves and the hardness ratio (i.e. count ratio in the
energy bands $0.3$-$1$ and $1$-$8$ keV bands) for two sources that show rapid
aperiodic variability. The spectral properties of these sources are tabulated in
Table $2$. The source in NGC 0628 is a ULX and its
aperiodic variability makes it unique in the sample.  }
\label{LC_var}
\end{minipage}
\end{figure}

\begin{figure}
\centering
\includegraphics[height=100mm, angle=0]{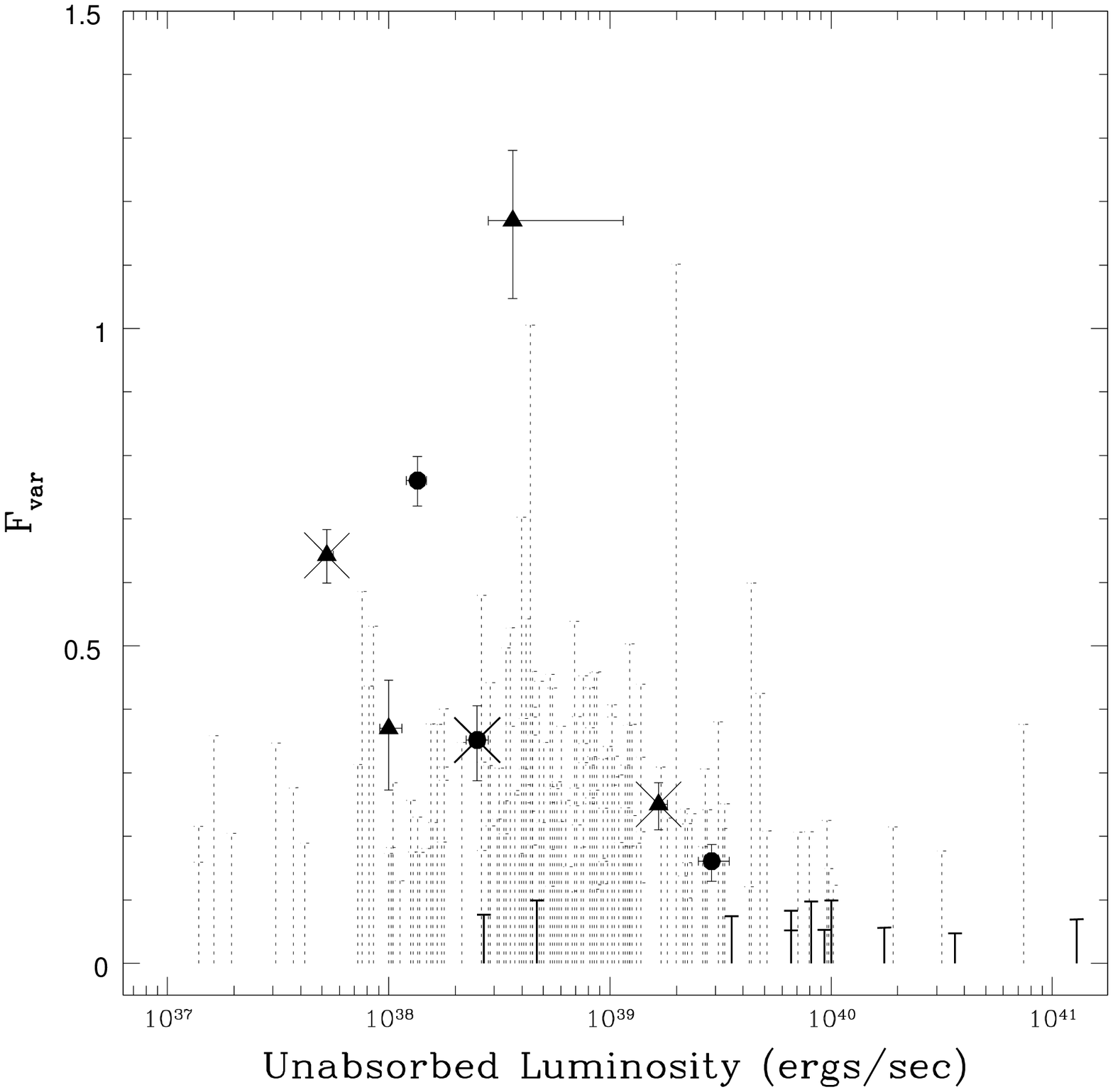}
\caption{Fractional root mean square variability  amplitude ($F_{var}$) versus the unabsorbed 
X-ray luminosity. The circles (triangles) indicate sources which are better fitted with a disk black body 
(power-law) spectral model. The sources marked with a star show rapid aperiodic variability. 
For 11 sources the 2-sigma  upper limits are constrained
to be less than 0.1 and these are represented by solid lines. Note that 9 of these 11 non-variable sources are ULXs (i.e. $L_X > 10^{39}$ erg s$^{-1}$
).
 }
\label{Fvar_lum}
\end{figure}

 Of the 166 sources, variability was detected
above the 3-sigma level for 5 sources and above the 2-sigma level for 8 sources.
One of these highly variable sources is a foreground star CXOU$~$J013647.4 +154745 \citep{Soria04}. We have not included
this source for further analysis. We are left with all total of 165 X-ray point sources out of which seven
are highly variable sources with
$F_{var}$ ranging from $0.16$-$1.16$. Their spectral parameters for
the absorbed power-law  and  disk black body models are tabulated in 
Table $2$. If all the 165 sources were not variable, one may expect by chance to get $\sim 4$ 
sources to show variability at a 2-sigma level. Hence, although we present the results of the two sources
that are variable at levels less than 3-sigma, we caution that some  of them may be spurious.
For most of the sources, $F_{var}$ is not well constrained being typically $F_{var} < 0.4$. 
However, for 11 sources we do obtain upper limits on $F_{var} < 0.1$.

The analysis is sensitive only to  large amplitude variations $F_{var} \ga 0.4$.
Such large variations are expected to be due to secular (i.e one-time) transitions
between flux levels. This is indeed true for four of the sources with detected variability.
The light curves for these are shown in Figure \ref{LC_sec}, where it can be seen that
the variability is due to secular transition  behaviour.  
The hardness ratio (HR) defined as the ratio of the counts in  $1$-$8$ and $0.3$-$1$  keV
bands, do not show such significant variation (Figure  \ref{LC_sec}). Thus, the 
large amplitude flux variations in these sources are not necessarily accompanied with
significant spectral shape changes. This implies that the variability of these sources is not due
to variable absorption as reported for a source in NGC 4472 \citep{Mac07}.  The large flux variation for the source
in NGC 1569 (S2 in Table $2$. ) 
has been reported earlier by \cite{Hei03}. The source is well
fitted by disk black body emission with bolometric luminosity varying from  $10^{37}$ erg s$^{-1}$
to $5 \times 10^{37}$ erg s$^{-1}$ suggestive of it being a stellar mass black hole system \citep{Hei03}.

\begin{table*}
\begin{minipage}{110mm}
\begin{center}
\caption{\small{Spectral Properties of variable sources} }
~\\
\tiny
\begin{tabular}{lcccccccccc}

\hline

Galaxy & Source & R.A & Decl. & {$(n_H)_{pow}$} & $\Gamma$ & {$log(L)_{pow}$}  & {$(n_H)_{dbb}$} & $kT_{in}$ & {$log(L)_{dbb}$ }  &{$F_{var}$} \\
       & No.&\\

\hline

NGC 1399 & S1 & 3 38 31.73 & -35 30 58.65 & $ 0.01^{+ 0.16}_{- 0.01}$ & $ 2.5^{+1.7}_{-0.6}$ & $38.5^{+0.5}_{-0.1}$ & $  0.00^{+ 0.06}_{-0.00}$ & $ 0.25^{+0.06}_{-0.06}$ & $38.82^{+0.26}_{-0.02}$  & $1.16^{+0.12}_{-0.12}$ \\
NGC 1569 & S2 & 4 30 48.14 & +64 50 50.57 & $ 0.51^{+ 0.06}_{- 0.06}$ & $ 4.0^{+0.3}_{-0.2}$ & $38.7^{+0.2}_{-0.2}$ & $  0.19^{+ 0.02}_{-0.03}$ & $ 0.41^{+0.04}_{-0.03}$ & $38.13^{+0.04}_{-0.05}$ & $ 0.76^{+0.03}_{-0.04}$ \\
NGC 1569 & S3 & 4 30 48.62 & +64 50 58.59 & $ 0.24^{+ 0.06}_{- 0.05}$ & $ 1.6^{+0.2}_{-0.1}$ & $37.72^{+0.03}_{-0.02}$ &  $  0.10^{+ 0.04}_{-0.05}$ & $ 1.68^{+0.30}_{-0.19}$ & $37.96^{+0.04}_{-0.04}$ &  $0.64^{+0.04}_{-0.05}$ \\
NGC 2403 & S4 & 7 37 11.63 & +65 33 45.52 & $ 0.63^{+ 0.26}_{- 0.18}$ & $ 2.3^{+0.4}_{-0.3}$ & $38.4^{+0.2}_{-0.1}$ & $  0.28^{+ 0.16}_{-0.14}$ & $ 1.09^{+0.23}_{-0.15}$ & $38.40^{+0.05}_{-0.05}$ &  $0.35^{+0.05}_{-0.06}$ \\
NGC 0628 & S5 & 1 36 51.06 & +15 45 46.86 & $ 0.03^{+ 0.05}_{- 0.03}$ & $ 1.9^{+0.2}_{-0.1}$ & $39.22^{+0.04}_{-0.02}$ & $  0.00^{+ 0.00}_{-0.00}$ & $ 0.89^{+0.08}_{-0.11}$ & $39.42^{+0.02}_{-0.04}$ & $0.25^{+0.03}_{-0.04}$ \\
\\

NGC 5128 & S6 &13 25 27.52 & -43  02 14.78 & $ 0.02^{+ 0.08}_{- 0.02}$ & $ 1.7^{+0.4}_{-0.3}$ & $38.0^{+0.06}_{-0.04}$ & $  0.00^{+ 0.10}_{-0.00}$ & $ 1.12^{+0.37}_{-0.23}$ & $38.20^{+0.08}_{-0.05}$ & $ 0.36^{+0.08}_{-0.09}$ \\
NGC 6946 & S7 &20 35  0.13 & +60  09  07.97 & $ 0.61^{+ 0.06}_{- 0.05}$ & $ 5.1^{+0.3}_{-0.3}$ & $40.4^{+0.2}_{-0.2}$ & $  0.22^{+ 0.03}_{-0.03}$ & $ 0.32^{+0.02}_{-0.03}$ & $39.46^{+0.08}_{-0.06}$  & $ 0.16^{+0.02}_{-0.04}$ \\

\hline
\end{tabular}
\end{center}
\tablecomments{1.45\textwidth}{ Host galaxy name; Source Name; Right Ascension; Declination; $n_H (\times 10^{22}$cm$^{-2}$), equivalent hydrogen column density; $\Gamma$,photon power-law index; L (erg s$^{-1}$ ), X-ray luminosity in the energy range: 0.3-8.0 keV; $kT_{in}$ (keV) is the inner disk temperature ; fractional r.m.s amplitude $F_{var}$. The error quoted on $F_{var}$ is 1-sigma.The first five sources are variable at greater the 3-sigma level while the rest are at greater the
2-sigma level. }
\label{table_podbb}
\end{minipage}
\end{table*}

A source in NGC 2403 (S4) showed a secular transition behaviour in the {\it Chandra} observation
analysed in the sample (Obs ID 2014). However, for three other {\it Chandra} observations, it
showed persistent large amplitude  variability. The light curves and hardness ratio for the
four observations are shown in Figure \ref{ngc2403}. The luminosity of this source
is $L \sim 10^{38}$ erg s$^{-1}$
 (S4 in Table $2$. ).
Two other sources exhibit aperiodic variability with large amplitudes and their
light curves are shown in Figure \ref{LC_var}. 
The source in NGC 1569 (like the one in NGC 2403) has a
relatively low luminosity of $L \sim 10^{38}$ erg s$^{-1}$
 (S3 in Table $2$. )
and its variable nature has been noted earlier by \citet{Hei03}. The other source which is
in NGC 0628 (S5 in Table $2$. ) has a luminosity $L > 10^{39}$ erg s$^{-1}$
 and 
is  unique  in the sample.
It shows large amplitude variations on time-scale as short
as $\leq 1000$ s. 
This source (also called M74 X-1) has been extensively studied by \citet{Kra05} using both {\it Chandra} 
and {\it XMM Newton} data. They have compared the variability of the source as being similar to that of the Galactic 
X-ray binary system, GRS 1915+105. Here we point out the rarity of such sources, since only one such object was
found in our sample which may be quantified as one object per 17 galaxies.

The high amplitude variability of the four sources shown in Figure \ref{LC_sec} may be  due to rare transitions/flares.
In that case, other observations of these sources may not show such variability. The source S6 of NGC 5128 does not show variability ($F_{var} \la 0.2$)
in seven other {\it Chandra} observations (IDs 2978, 3965, 7797, 7798, 8489, 8490, 10722).
Unfortunately,
there are no additional {\it Chandra} observations of NGC 1569 (with an exposure time $> 30$ ksec) to confirm the
nature of the sources S3 or S2.

Figure \ref{Fvar_lum} shows the fractional r.m.s variability amplitude, $F_{var}$ versus the
the unabsorbed X-ray luminosity. For the seven sources with detected variability, the
triangles represent sources whose  spectra are better 
modelled as a power-law, while the circles are for those with spectra
better modelled as a disk black body. The three sources marked with crosses exhibit aperiodic
fluctuations (Figure \ref{ngc2403} and Figure \ref{LC_var} ) in contrast to the other variable sources which
exhibit secular transitions/flares (Figure \ref{LC_sec}). One of the ULXs has a ultra-soft spectrum with disk black body 
temperature $\sim 0.3 $ keV. The other is the rapidly varying source in NGC 0628.
For non-variable sources, the upper limits
are plotted in the Figure. For 11 sources the upper limits are constrained
to be less than 0.1 and these are represented by solid lines and nine of them have luminosities
in excess of $L > 10^{39}$ erg s$^{-1}$ .  
Therefore, it seems that  ultra-luminous X-ray sources are typically not highly variable in these 
kilo-second time scale except for some ultra-soft one.

\section{Discussion}

Of the 165 ( not including foreground star) sources in 17 nearby galaxies, seven of them were found to be highly variable in k-sec time-scales.
In general, the variability is not strongly correlated with hardness ratio, which indicates that they are
not due to varying absorption, instead they are likely to represent some structural changes in the accretion process.

In the sample, we find that while there are two ULXs which are variable ($F_{var} > 0.2$), there are
nine ULXs for which we can constrain the variability $F_{var} < 0.1$. One of the variable ULXs is ultra-soft 
with black body temperature $ \sim 0.3$ keV. 
For such sources, the spectra in {\it Chandra} energy range of $0.3$-$8.0$ keV
is essentially an exponentially decreasing function. Relatively small fluctuations
in an intrinsic spectral parameter, may lead to large flux variation in this spectral
regime. Thus it seems  ULXs are not highly variable in ksec time-scales.
A notable exception is the bright source in NGC 0628 whose variability is unique
in the sample.

For four of the seven variable sources (i.e. $\sim 3$\% of the sample), the variability seems to be of a secular nature, 
perhaps representing state transitions.
At least one of them, is not variable in seven other {\it Chandra} observations, which implies that
the variability is of a secular nature and not persistent. 
The analysis is based on single ``snapshot'' observation of $\sim 40$ ksec.
If one hypotheses, that all sources undergo such transitions, the detection of such variability in $\sim 3$\% of the sources implies
that on the average the duty cycle of such transitions  is $\sim 40000/0.03 \sim 1.3 \times 10^{6}$ secs or a 
transition every $\sim 15$ days. A similar analysis for ULXs where there is one source showing a secular variability for
nine sources that don't would imply a transition every $\sim 40000/0.1 \sim 4 \times 10^{6}$ secs $\sim 5$ days. 
However, since the sample considered here is neither uniform or complete, such
quantitative statistical inference may not be reliable.
Nevertheless, since spectral state transitions for Galactic sources typically occur much less frequently, it is more likely that these
sources represent a special class of systems where transitions occurring on k-sec time-scales, happen frequently.
 Further, state transitions in Galactic sources are associated with spectral changes, while the hardness ratio
does not change significantly during these transitions. Thus, it seems these transitions are different from the state
transitions observed in  Galactic sources.

For three of the variable sources the large amplitude fluctuations are persistent. Among Galactic X-ray binaries,
GRS 1915+105 and the recently discovered IGR J17091-3624 \citep{Alt11,Pah12} show such persistent variability in these time-scales. Two of these sources, 
unlike GRS 1915+105 but perhaps similar to IGR J17091-3624, have relatively low luminosity of $\sim 10^{38}$ erg/s. A source
in NGC 2403 shows large amplitude variability in all four {\it Chandra} observations. It is possible, that the
variability maybe quasi-periodic in  $\sim 10$ ksec time-scale, but the data is not sufficient to make
concrete statements. Thus, we establish that there are  X-ray sources in nearby galaxies, with modest luminosities
of $L \sim 10^{38}$ erg s$^{-1}$ , that exhibit persistent high amplitude variability in  ksec time-scales.

It is premature to identify any specific accretion disk instability model to explain these
large amplitude variability. Indeed, the variability of the Galactic source GRS 1915+105 is
still not clearly understood. Nevertheless, with the increased number of sources, detailed studies like
flux resolved spectral analysis and time-lag analysis may shed light on the reasons
for their variable nature.

\section{Acknowledgements}
This research has made use of data obtained from the High Energy Astrophysics Science Archive Research Center (HEASARC), provided by NASA's Goddard Space Flight Center and software provided by Chandra X-ray Center (CXC) in the application packages and tools. SM gratefully acknowledges IUCAA for the visiting associateship.

\end{document}